# A Service-Based Approach for Managing Mammography Data

**Florida Estrella[a,b], Richard McClatchey[a], Dmitry Rogulin[a,b], Roberto Amendolia[b], Tony Solomonides[a]**

*[a] Centre for Complex Cooperative Systems, University of the West of England, Frenchay, Bristol , UK*
*[b] ETT Division, CERN, 1211 Geneva 23, Switzerland*

**Abstract**

*Grid-based technologies are emerging as a potential open-source standards-based solution for managing and collaborating distributed resources. In view of these new computing solutions, the Mammogrid project is developing a service-based and Grid-aware application which manages a European-wide database of mammograms. Medical conditions such as breast cancer, and mammograms as images, are extremely complex with many dimensions of variability across the population. An effective solution for the management of disparate mammogram data sources is a federation of autonomous multi-centre sites which transcends national boundaries. The Mammogrid solution utilizes the Grid technologies to integrate geographically distributed data sets. The Mammogrid application will explore the potential of the Grid to support effective co-working among radiologists throughout the EU. This paper outlines the Mammogrid service-based approach in managing a federation of grid-connected mammography databases.*

## Introduction

The medical community has been exploring collaborative approaches to the sharing and analysis of mammography data. The next major advances in the evidence base through the study of mammograms will follow from interaction in an environment that transcends geography and culture. The National Digital Mammography Archive (NDMA) project [1] is developing a national breast imaging archive and network infrastructure to support digital mammography using next generation internet technologies. The eDiamond project [2] aims to develop a grid-based system which ensures database consistency and reliable image processing through image standardization.

Complementary to these efforts, the MammoGrid (MG) project [3] aims to prove that Grids infrastructure can be practically used for collaborative medical image analysis. One of the main deliverables of this project is a demonstrator software using open-source Grid middleware that is capable of managing federated mammogram databases distributed across Europe. The proposed solution is a service-based grid-aware framework encompassing regions of varying protocols, lifestyles and diagnostic procedures.

The framework will allow, among other things,

1. mammogram data mining,
2. diverse and complex epidemiological studies,
3. statistical and computer aided detection (CADe) analyses &
4. deployment of versions of the standardization system Standard Mammogram Form (SMF).

SMF [4] enables comparison of mammograms in terms of tissue properties independently of scanner settings.

This paper discusses a service-based approach for managing mammography data. The proposed solution is a vertical stack of grid services which are coordinated to handle distributed mammograms. The distributed mammogram sources are federated into a single virtual organization.

## Underpinning Technologies

The adopted Grid implementation is ALICE Environment (AliEn). AliEn [5] is an open-source Grid implementation developed to satisfy the needs of the ALICE experiment at CERN for large scale distributed computing. It provides the essential middleware services that allow access to distributed resources in an AliEn-connected environment.

Some of the MG-relevant AliEn services include

1. authentication for checking user credentials,
2. resource broker for job and algorithm scheduling,
3. storage element for data and file management, &
4. file transfer for scheduled file transfer functionality.

[1] Keywords should, preferably, be drawn from the MeSH thesaurus

There are many concepts behind this Grid philosophy, and these are discussed in the following sections.

**Service-Oriented Architecture**

A *service* is an entity that provides some capability through exchange of messages. A Service-Oriented Architecture (SOA) [6] is a collection of interconnected communicating services. The philosophy underpinning the SOA is that services are described through a service description language, that services are dynamically discovered by applications, and that service methods are invoked through the communication protocol defined in the service interface.

A typical SOA behaviour is as follows:

1. a service is created
2. the service is published
3. the service is located
4. the service is invoked
5. optionally, the service is unpublished

The technology of Web Services (WS) [7, 8] is the most likely service connection technology for SOA. WSs are self-contained, modular applications that can be described, published, located and invoked over a network. The WS architecture allows for loosely coupled dynamically bound components. Figure 1 [9] illustrates the main roles (*service provider*, *service requester*, *service broker*) and operations (*publish*, *locate*, *bind*) in a WS architecture.

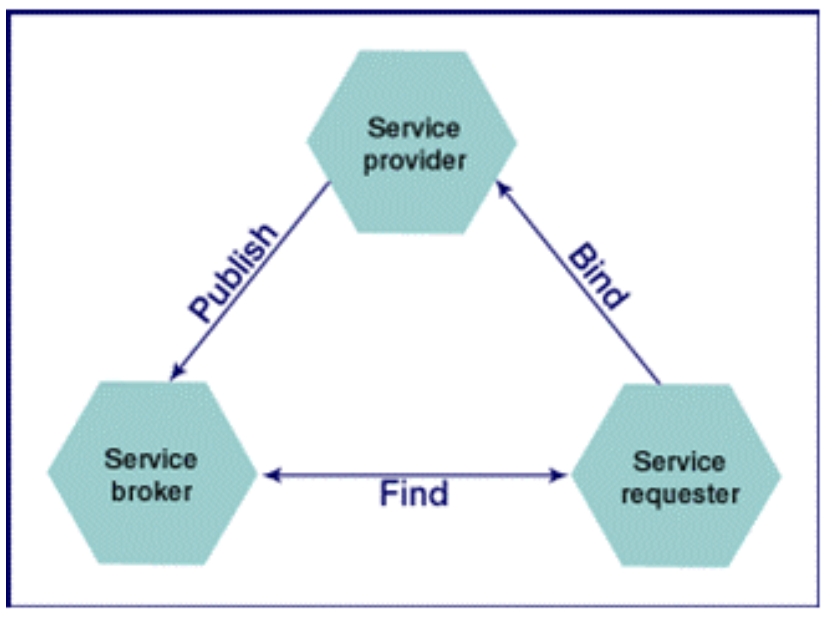

*Figure 1- Service Oriented Architecture*

**The Grid**

A virtual organization (VO) [9] is a flexible, secure, coordinated resource sharing among dynamic collections of individuals, resources and institutions. A VO scenario – thousands of physicists, laboratories and institutes across the world collaborating to study and analyze physics data coming from the CERN detector [10]. Disparate computing, storage and network resources are orchestrated to achieve a shared physics goal.

The Grid is a SOA that supports the sharing and coordinated use of diverse resources in a VO. The Open Grid Service Architecture (OGSA) is a standard specification for grid-based applications. OGSA defines what grid services (GS) are, what they should be capable of, what types of technologies they should be based on. The Open Grid Service Infrastructure (OGSI) [11] is a formal and technical specification of the concepts described in OGSA. The Globus Toolkit 3 (GT3) [12] is currently the only usable implementation of OGSI. These concepts are illustrated in Figure 2.

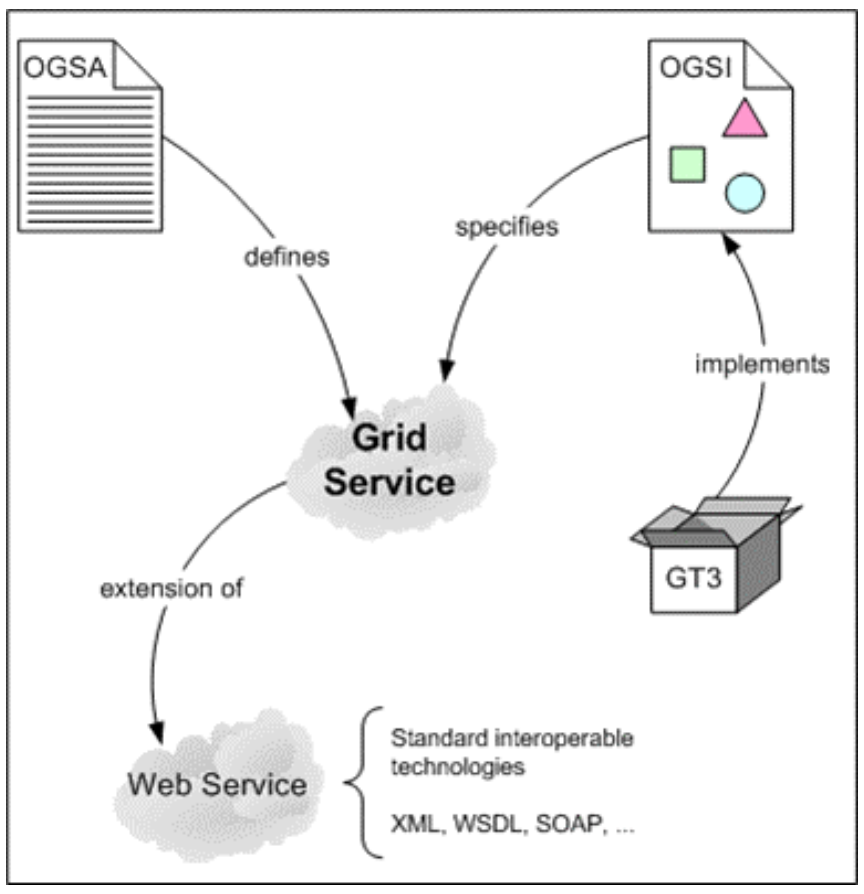

*Figure 2 - Grid Services*

WSs are stateless, where data are not retained among invocations. OGSA defines GSs as stateful extensions of WSs. A chain of coordinated invocations cannot be handled by 'normal' WSs. One good example is complex physics simulation which normally requires a sequence of inter-related actions.

WSs are persistent, where services outlive their clients. GSs are time-limited services with delegated authentication credentials. The mammography domain, and similar areas dealing with sensitive and confidential data, requires services which are capable of dealing with authentication certificates.

OGSA draws on the same infrastructure as in WSs – HTTP, XML, SOAP, WSDL.

| | |
|---|---|
| HTTP | http://www.w3.org/Protocols/rfc2068/rfc2068 |
| XML | http://www.w3.org/XML/ |
| SOAP | http://www.w3.org/TR/SOAP/ |
| WSDL | http://www.w3.org/TR/wsdl |
| UDDI | http://www.uddi.org/ |

The next section discusses how these underlying technologies can be applied to medical digital images.

## The Mammogrid Solution

Medical conditions such as breast cancer, and mammograms as images, are extremely complex with many dimensions of variability across the population. Similarly, the way diagnostic systems are used and maintained by clinicians varies between imaging centres and breast screening programmes, and in consequence so does the appearance of the mammograms generated. An effective solution for the management of disparate mammogram data sources is a federation of autonomous multi-centre sites which transcends national boundaries.

The Mammogrid solution utilizes the Grid technology to enable distributed computing at a European scale. Figure 3 illustrates this configuration.

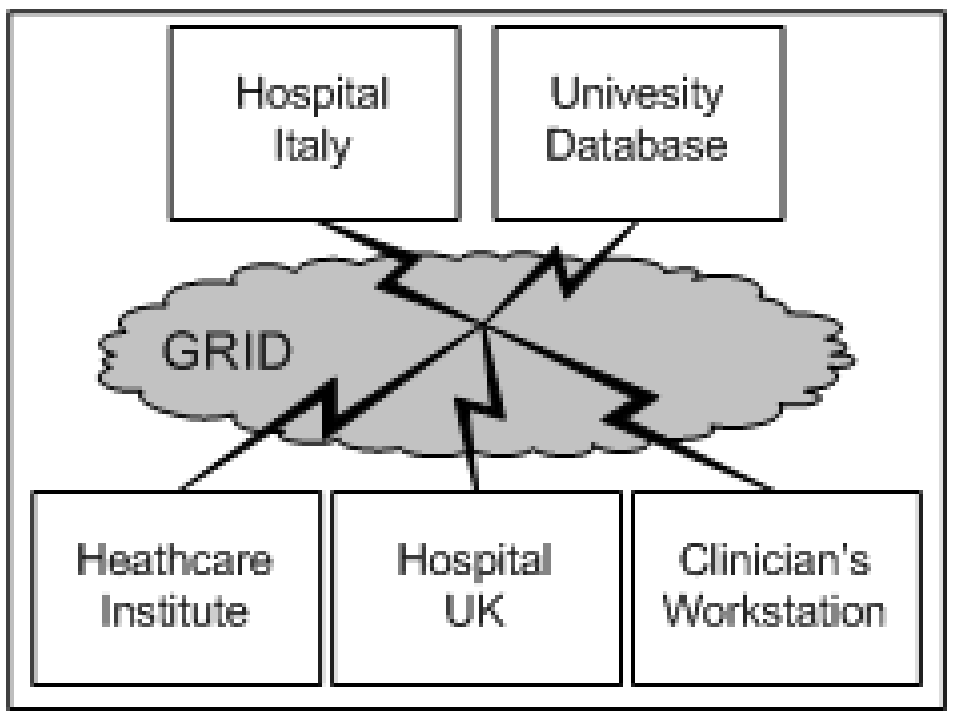

*Figure 3 Distributed Mammogram Data*

There are two federation scenarios –

1. a federation in a single VO and
2. a federation of multi-VOs.

These are discussed below.

**Federation in a Single Virtual Organisation**

The resources in the federation, e.g. hospitals, research institutes, universities, are governed by the same sharing rules with respect to authentication, authorization, resource and data access. These rules create a highly controlled environment which dictates what is shared, who is allowed to share, and the conditions under which sharing occurs among medical sites. Federation in this application, implies cooperation of inpendent medical sites. Individually, these sites are autonomous in that they have separate and independent control of their local data. Collectively, these sites participate in a federation, and the federation is governed by the organization.

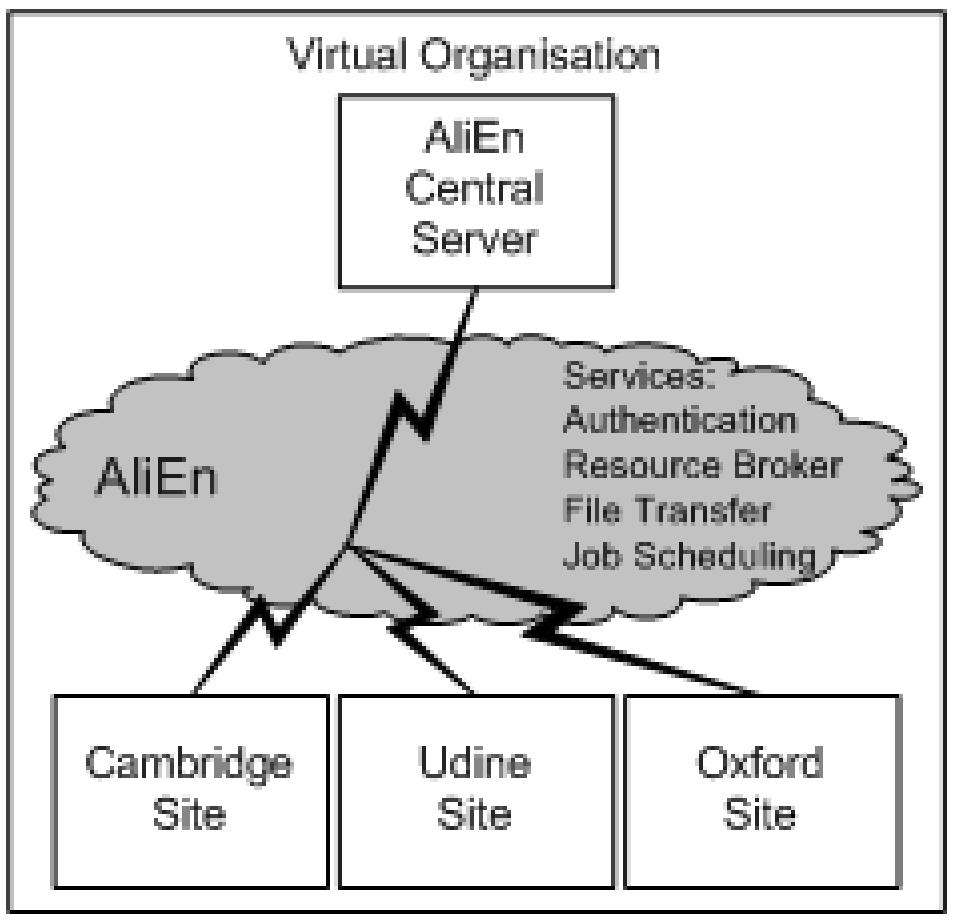

*Figure 4 Federation in an AliEn-connected VO*

Figure 4 illustrates an AliEn-connected single VO configuration. The AliEn middleware provides services (e.g. authentication, data access, resource broker, file transfer) that facilitate the management of resources in the VO. In essence, the medical community dictates the interaction protocol, and AliEn implements and enforces these rules on the participating entities of the organization through services. For detailed discussion, see the Mammogrid Technical Specification in [3].

**Federation of Multiple VOs**

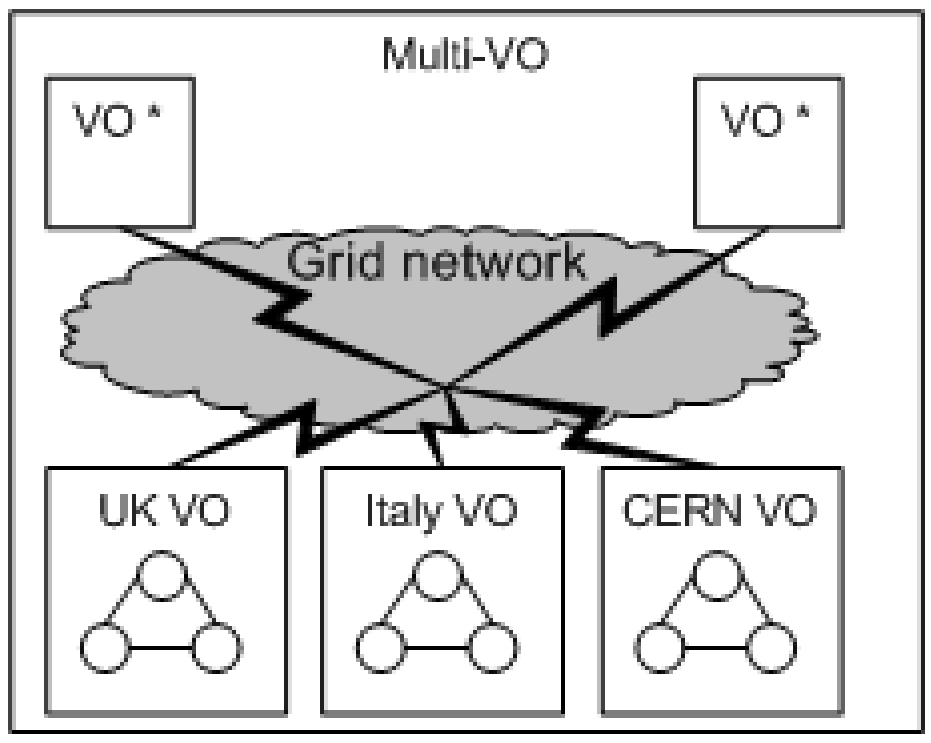

*Figure 5 Multi-VO Configuration*

The medical sites in a single VO operate within the rules specified by a governing organization. In reality, there are many (co)-existing organizations, with different rules and protocols. Typically, hospitals have different regulations and governments have different legislations. A federation of multiple VOs extends the single VO setup by inter-connecting potentially disparate VOs. This configuration is beyond the scope of this paper. See Figure 5.

**Vertical Stack of Grid Services**

Medical image management on the Grid uses a vertical stack of grid services (from here on, referred to as services). This is illustrated in Figure 6. The services are vertically layered that build open one another. The services are 'orchestrated' in terms of service interactions – how services are discovered, how services are invoked, what can be invoked, the sequence of service invocations, and who can execute.

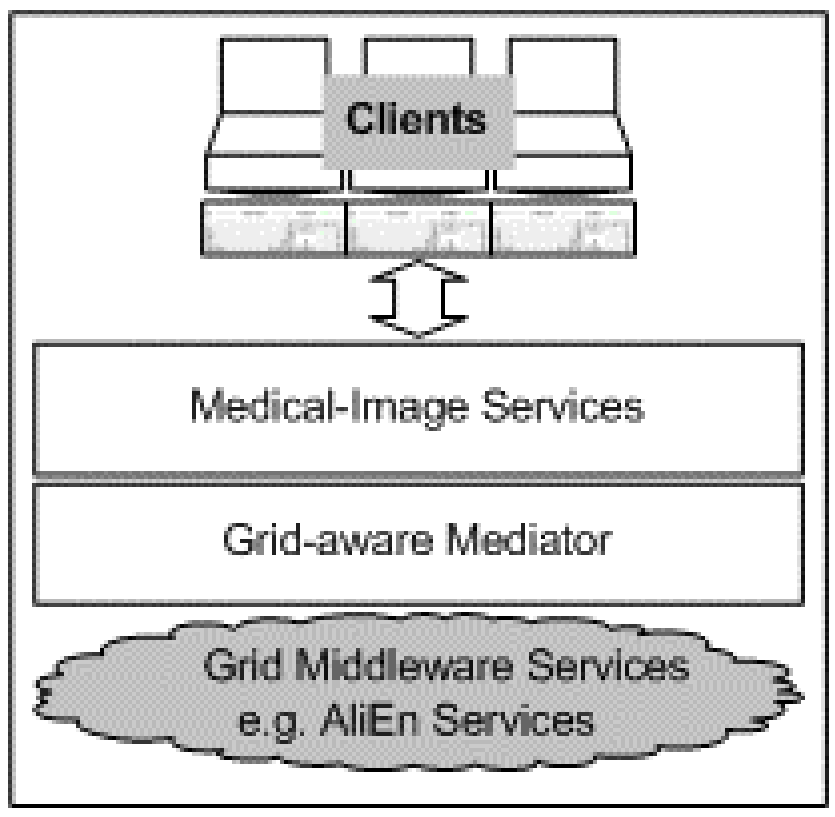

*Figure 6 Vertical Stack of Grid Services*

The medical images (MI) services are directly invoked by authenticated MG clients. They provide a generic framework for managing image and patient data. The digitized images are imported and stored in DICOM [13] format. Relevant MI services include:

1. *add* for uploading DICOM files
2. *retrieve* for downloading DICOM files
3. *update* for updating contents of DICOM files

4. *query* for translating and executing a query on patient data
5. *addAlgorithm* for loading a executable code
6. *executeAlgorithm* for executing an algorithm
7. *authenticate* for logging into the system.

It is intended that versions of the SMF and CADe algorithms will be uploaded and executed on grid-resident images. This supports grid-enabled monitoring of the quality of mammograms submitted at each participating centre.

The Grid-aware (GA) mediators are arbitrating components between MI-based services and the underlying OGSA-compliant Grid middleware. These services control and coordinate the hospital interactions with the encoded knowledge of the MI-domain and the Grid. As mediators, these services promote loose coupling between the MI-domain and the Grid-middleware. Some important GA components -

1. *DataManager* for storing and retrieving DICOM files on the Grid
2. *QueryManager* for executing queries on the Grid

The Grid middleware services address issues related to resource management, information discovery and security infrastructure. These services are expected to be OGSA compliant thus paving the way for interoperability. The OGSA-compliant GT3 includes GRAM, MDS-2 and GSI. The Grid Resource Allocation and Management (GRAM) protocol provides for secure and reliable service creation and management. The Meta Directory Service (MDS-2) deals with information discovery, data modeling and registry. The Grid Security Infrastructure (GSI) supports single sign on, delegation and credential mapping.

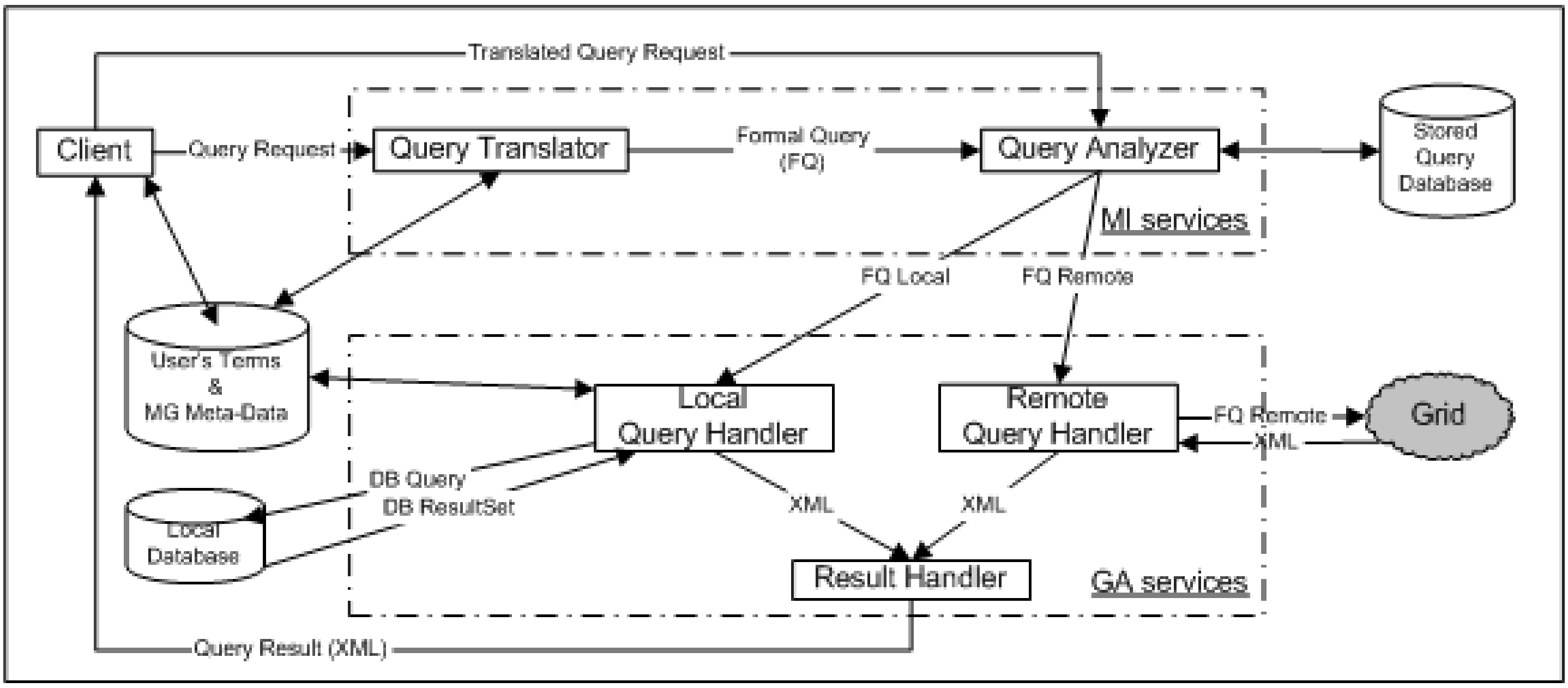

*Figure 7 Query Handling in Mammogrid*

**Query Handling in Mammogrid**

The MG demonstrator aims to facilitate mammogram analysis in terms of 'real clinician queries' against Grid-resident images. Simple queries like '*find all mammographic images for all women over 50 undergoing HRT treatment*', complex queries like '*find all patients who have developed cancer in the other breast after successful therapy on the first cancer*', and queries related to CADe or SMF results are expected to be formulated and executed by radiologists on potentially distributed images, just as if these images are locally resident.

Figure 7 illustrates how queries are handled – from formulation to result – using a stack of orchestrated services. The sequence of events is as follows:

(1) Clients (e.g. end-users, applications) define their mammogram analysis in terms of queries they wish to be resolved across the collection of data repositories (either locally- or remotely-held data). This uses descriptive information (User's Terms and MG-specific metadata) about the query domain (both graphical specifications and user-specific terms) to translate the user query into a 'data request' using standard terms.

(2) Query Translator takes the user request and translates to a MG-defined formal query representation.

(3) Queries are executed at the location where the relevant data resides. That is, the sub-queries are moved to the data, rather than large quantities of data being moved to the clinician, which is prohibitively expensive given the quantities of data. The Query Analyser takes a formal query representation and de-composes into (a) formal query for local processing and (b) formal query for remote processing. It then forwards these de-composed queries to the Local Query Handler and the Remote Query Handler for the resolution of the request.

(4) The Local Query Handler generates query language statements (e.g. SQL) in the query language of the associated Local DB (e.g. MySQL). The result set is converted to XML and routed to the Result Handler.

(5) The Remote Query Handler is a portal for propagating a queries and results between sites. This handler forwards the formal query for remote processing (3b above) to the Query Analyser of the remote site. The remote query result set is converted to XML and routed to the Result Handler.

(6) The Result Handler is responsible for collecting query results – both local and remote. The query handlers return XML results, and these are "joined" to create the overall result to be sent back to the requestor – either the client of the Remote Query Handler.

Medical imaging services include

1. *QueryTranslator* for translating user defined queries to a MG-defined formal query representation and
2. *QueryAnalyzer* decomposes a formal query for local processing and remote processing.

Whereas Grid-aware services include the

1. *LocalQueryHandler* generates query language statements (e.g. SQL) for the local database (e.g. MySQL)
2. *RemoteQueryHandler* is a portal for propagating queries and results between grid sites.
3. *ResultHandler* collects query results, local and remote, and returns XML to the requestor.

**Conclusions**

The MammoGrid project has recently delivered its first proof-of-concept prototype enabling clinicians to store (digital) mammograms along with appropriately anonymised patient meta-data and to provide controlled access to mammograms both locally and remotely stored. A typical database comprising several hundred mammograms is being created for user tests of the query handler. The prototype comprises a high-quality clinician visualization workstation used for data acquistion and inspection, a DICOM-compliant interface to a set of medical services (annotation, security, image analysis, data storage and querying services) residing on a so-called 'Grid-box' and secure access to a network of other Grid-boxes connected through Grids middleware. Clinicians are being closely involved with these tests and it is intended that a subset of the clinician queries listed in section 3 will be executed to solicit user feedback. Within the next year a rigorous evaluation of the prototype will then indicate the usefulness of the Grid as a platform for distributed mammogram analysis and in particular for resolving clinicans' queries. The system will be tuned for performance and for security prior to the release of a second prototype at the end of the project in mid 2005. It is intended that the MammoGrid medical services for this second prototype will adhere to emerging Grids standards e.g [11].

The Grid platform provides an effective distributed computing model for harnessing the use-of and access-to massive amounts of medical image data across Europe. Moreover, it will enable a standardized, distributed digital mammography resource for improving diagnostic confidence

The use of web services or grid services allows for containment and modularity, minimizing the requirements for shared understanding. This platform creates a dynamic loosely coupled client-server environment. That services can be described, published, located and invoked over the WWW is a plus to an already very promising framework.

**Acknowledgements**

The authors thank their home institutes and acknowledge the support of the following Mammogrid Collaboration members. UWE: Hauer T, Manset D, Odeh M; CERN: Galvez J, Buncic P, Saiz P; Mirada: Schottlander D, Reading T, Highnam R.

**Address for correspondence & Lead author:**

Dr Florida Estrella, ETT Division, CERN
1211 Geneva 23, Switzerland